\documentclass[12pt,a4paper]{article}
\usepackage{a4wide}
\usepackage{amsmath}
\usepackage{latexsym}
\begin{document}
\def\d{{\rm d}}
\def\reg{{\rm reg}}

\def\veck{{\pmb{k}}}
\def\vecp{{\pmb{p}}}
\def\vecv{{\pmb{v}}}

\def\vecE{{\pmb{E}}}
\def\vecF{{\pmb{F}}}
\def\vecH{{\pmb{H}}}

\def\vecalpha{{\pmb{\alpha}}}
\def\vecgamma{{\pmb{\gamma}}}
\def\vecvarepsilon{{\pmb{\varepsilon}}}
\def\vecnabla{{\pmb{\nabla}}}

\newcommand{\Atop}[2]{\genfrac{}{}{0pt}{}{#1}{#2}}

\pagenumbering{arabic}

\begin{center}
\begin{titlepage}
\vspace*{2cm}

{\bf\large The Effective Lagrangian of QED with a Magnetic Charge}
\vspace{2.0cm}

{\sc\large V. G. Kovalevich}

\vspace{0.5cm}

{\it  Institute of Physics, 220072 Minsk, Belarus}
\vspace*{1cm}

{\sc \large P. Osland}
\vspace{0.5cm}

{\it Department of Physics, University of Bergen, 
    All\'egt.~55, N-5007 Bergen, Norway}
\vspace{1cm}

{\sc\large  Ya.\ M. Shnir}
\vspace{0.5cm}

{\it Department of Mathematics, 
Technical University of Berlin, 10623 Berlin, Germany }{\footnote{
Supported by the Alexander von Humboldt
Foundation; permanent address:
Institute of Physics, 220072 Minsk,
Belarus}}

\vspace{1cm}

{\sc\large E. A. Tolkachev}
\vspace{0.5cm}

{\it  Institute of Physics, 220072 Minsk, Belarus}
\vspace*{1cm}

  \begin{abstract}
The effective Lagrangian of QED coupled to dyons is calculated. 
The resulting generalization of the Euler-Heisenberg Lagrangian
contains non-linear $P$- and $T$-nonivariant (but $C$ invariant) 
terms corresponding to the virtual pair creation of dyons. 
As examples, the amplitudes for photon splitting 
and photon coalescence are calculated. 

 \end{abstract}

\end{titlepage}
\end{center}

The calculation of quantum corrections due to the virtual 
pair creation of dyons is a very difficult problem 
because the standard 
diagram technique is not valid in this case. 
The difficulty is connected both to the large value 
of the magnetic charge of the dyon and the lack of a
consistent local Lagrangian formulation of electrodynamics with two
types of charge \cite{StrTom}. 
So, there is no possibility to use a perturbation expansion
in a coupling constant. 
But one can apply the loop expansion which is just an expansion 
in powers of the Planck constant $\hbar$.

Actually, it is known (see, e.g. \cite{AkhiBer}),
that the one-loop quantum correction to the QED Lagrangian 
can be calculated without the use of perturbation methods. 
The correction is just the change in the vacuum energy
in an external field. 
Let us review the simple case of weak constant parallel electric and
magnetic fields $\vecE$ and $\vecH$. We impose the conditions
$e |\vecE| /m^2 \ll 1$ and $e |\vecH| /m^2 \ll 1$ such that the creation
of particles is not possible. 
In this case the one-loop correction can be calculated 
by summing the one-particle modes --- the solutions of the Dirac
equation in the external electromagnetic field --- over all quantum
numbers \cite{AkhiBer}, \cite{Wentzel}. For example, if there is just
a magnetic field, $\vecH = (0, 0, H)$, 
the corresponding equation is 
\begin{equation}                                    \label{Dirac}
[i\gamma^{\mu}(\partial_{\mu} + i e A_{\mu}) - m ] \psi (x) = 0
\end{equation}
where the electromagnetic potential is
$A^{\mu} = (0, - Hy, 0, 0)$. 
The solution of this equation gives the energy
levels of an electron in a magnetic field \cite{BerLifPit}, \cite{Bagrov}
\begin{equation}                                     \label{ELand}
\varepsilon _n = \sqrt{m^2 + e H (2n - 1 + s ) + k^2}
\end{equation}
where $n = 0, 1, 2 \dots $, $s = \pm 1$, and $k$ is the electron 
momentum along the field.
In this case the correction to the Lagrangian is 
\cite{AkhiBer}, \cite{BerLifPit} 
\begin{eqnarray}                                      \label{DelE}
\Delta L_H = \frac{e H }{2 \pi ^2} \int \limits _0^{\infty} \d k
\left[ (m^2 + k^2) ^{1/2} 
+ 2 \sum _{n= 1} ^{\infty} (m^2 + 2eH n + k^2 )^{1/2}\right] 
\nonumber      \\
= - \frac{1}{8 \pi ^2} \int \limits _0^{\infty} \frac{\d s}{s^3} 
e^{-m^2 s} \left[
(esH)~{\rm coth}~(esH) - 1 - \frac{1}{3} e^2 s^2 H^2 \right],
\end{eqnarray}
where the terms independent of the external field
$\vecH$ are dropped and a standard renormalization 
of the electron charge has been made \cite{AkhiBer}.

It is known \cite{Bagrov} that if we consider simultaneously
magnetic ($\vecH$) and electric ($\vecE$)
homogeneous fields, then equation (\ref{Dirac}), 
as well as its classical analogue can be separated into 
two uncoupled equations, each in two variables.
Indeed, in this case we can take $A^{\mu} = (Ez, - Hy, 0, 0)$ 
and the interaction of an electron with the fields
$\vecE$ and $\vecH$ determined independently. 
For such a configuration of electromagnetic fields 
the correction to the Lagrangian is
(see \cite{AkhiBer}, p.~787)
\begin{eqnarray}                                  \label{DelE_E}
\Delta L = \frac{e H}{2 \pi ^2} \sum _{n=1}^{\infty}
\int \limits _0^{\infty} \d k \, \varepsilon _n^{(E)} (k).
\end{eqnarray}
Here $\varepsilon _n^{(E)}$ is the correction to the energy of
an electron in the combined external magnetic and electric
fields, which is in the first order proportional to $e^2 E^2$.

So, the total Lagrangian is $L = L_0 + \Delta L$, where
$L_0 = (\vecE^2 - \vecH^2)/2$~~ 
is just the Lagrangian of the free electromagnetic field
in the tree approximation, and can be written as
\begin{eqnarray}                                  \label{Ltot}
L = \left( 1 + \frac{\alpha}{3 \pi} \int \limits _{0}^{\infty}
\frac{\d s}{s} e^{-m^2 s} \right) \frac{\vecE^2 - \vecH^2}{2} 
+ \Delta L' .
\end{eqnarray}
The logarithmic divergency can be removed by the standard renormalization
of the external fields and the electron charge:
\begin{equation}                                     \label{renorm}
E_{\reg} = Z_3^{-1/2} E; \qquad H_{\reg} = Z_3^{-1/2} H; \qquad
e_{\reg} = Z_3^{1/2} e,
\end{equation}
where $Z_3^{-1} = 1 + \frac{\alpha}{3 \pi} \int \limits _{0}^{\infty}
\frac{ds}{s} e^{-m^2 s}$ is just the usual QED renormalization factor.
Thus the finite part of the correction to the Lagrangian
$\Delta L'$ can be written in terms of physical quantities as 
(see \cite{AkhiBer}, p.~790)
\begin{eqnarray}                                              
\label{DeltaL}
\Delta L'
= - \frac{1}{8 \pi ^2} \int \limits _0^{\infty} \frac{\d s}{s^3} 
e^{-m^2 s} \left[
(esE) (esH)~{\rm cot}(esE) {\rm coth}(esH) - 1
\right],
\end{eqnarray}
which in the limit $E=0$ reduces to the renormalized form
of (\ref{DelE}).

The series expansion of (\ref{DeltaL})
in terms of the parameters  $e E /m^2 \ll 1$, $e H /m^2 \ll 1$ yields
the well known Euler-Heisenberg correction \cite{Eu-Hei}:
\begin{equation}                
\label{DeltaL-Euler}
\Delta L' \approx \frac{e^4}{360 \pi ^2 m^4} 
\Bigl[ (\vecH^2 - \vecE^2)^2 +
7 (\vecH \vecE)^2 \Bigr],
\end{equation}
where $e^2=\alpha$.{}\footnote{There is a misprint in 
ref.~\cite{AkhiBer}, they use $e^2=\alpha$ for this formula,
but elsewhere (p.~122) they have $e^2/4\pi=\alpha$.}

Let us consider how the situation changes if we consider
the virtual pair creation of dyons in the external electromagnetic field.
Using an analogy with the classical Lorentz force on a dyon 
of velocity $\vecv$
with electric ($Q$) and magnetic ($g$) charges \cite{StrTom}
\begin{equation}
\vecF 
=  Q\vecE + g \vecH + \vecv \times (Q\vecH - g\vecE),
\end{equation}
we shall assume that the wave equation for this particle in an external
electromagnetic field can be expressed as \cite{BlagSen}
\begin{equation}                                 \label{Dirac-dyon}
(i \gamma^{\mu} D_{\mu} - M) \psi (x) = 0,
\end{equation}
where $M$ is the dyon mass, and $iD_\mu$ 
a generalized momentum operator, with
$D_{\mu} = \partial _{\mu} + i Q A_{\mu} 
+ i g B_{\mu}$.{}\footnote{We would like to 
stress that Eq.~(\ref{Dirac-dyon}) is a postulate. 
For a discussion of the self-consistency of this approach, 
see, e.g. \cite{StrTom}, \cite{Zwanz}, \cite{BlagSen}. }

The potential $A_{\mu}$ and its dual $B_{\mu}$ are defined 
by $F_{\mu \nu} 
= \partial _{\mu} A_{\nu} - \partial _{\nu} A_{\mu} 
=\varepsilon _{\mu \nu \rho \sigma} \partial^{\rho} B^{\sigma}
$ where $ F_{\mu \nu} $ is the electromagnetic field strength  
tensor\footnote{ 
This definition is consistent only if $\Box 
A_{\mu} = \Box B_{\mu} = 0$, i.e., for constant electromagnetic fields 
or for free electromagnetic waves.}
and $\varepsilon_{0123}=1$. 
The potentials in the case of constant parallel electric and
magnetic fields can be expressed as
\begin{equation}                               \label{dual-pot}
A^{\mu} = (Ez, - Hy, 0, 0), \qquad B^{\mu} = (Hz, Ey, 0, 0).
\end{equation}

It is easily seen that the solution to the equation of motion for a dyon
in an external electromagnetic field can be obtained from the
solution to the equation for an electron (\ref{Dirac}) 
by the following substitution
\begin{equation}                               \label{anzatz}
e E \to Q E + g H; \qquad  e H \to Q H - g E.
\end{equation}
Using the same substitution as in Eqs.~(\ref{Ltot}) and (\ref{DeltaL}), 
we obtain the following expression for the quantum correction 
to the Lagrangian, due to the vacuum polarization caused by dyons:
\begin{eqnarray}                                 \label{Ltot-dual}
L = \left( 1 + \frac{Q^2}{12 \pi ^2} \int \limits _{0}^{\infty}
\frac{\d s}{s} e^{-M^2 s} 
- \frac{g^2}{12 \pi ^2} \int \limits _{0}^{\infty}
\frac{\d s}{s} e^{-M^2 s} \right)
\frac{\vecE^2 - \vecH^2}{2} 
+ \Delta L' ,
\end{eqnarray}
where a total derivative has been dropped.

For the renormalization of this expression we can introduce 
the renormalization factors \cite{BlagSen}
\begin{equation}                           \label{factrenorm-dual}
Z_e^{-1} = 1 + \frac{Q^2}{12 \pi} \int \limits _{0}^{\infty}
\frac{\d s}{s} e^{-M^2 s}; \quad
Z_g^{-1} = 1 - \frac{g^2}{12 \pi} \int \limits _{0}^{\infty}
\frac{\d s}{s} e^{-M^2 s},
\end{equation}
which are generalizations of the definition $Z_3$ of Eq.~(\ref{renorm}). 
In this case the fields and charges are renormalized as
\cite{BlagSen}
\begin{equation}                                               
\label{renorm-dual}
E_{\reg}^2 = Z_e^{-1}Z_g^{-1} E^2; \qquad H_{\reg}^2 
= Z_e^{-1}Z_g^{-1} H^2; \qquad
e_{\reg}^2 = Z_e Z_g e^2;\qquad g_{\reg}^2 = Z_e^{-1} Z_g^{-1} g^2.
\end{equation}
This relation (\ref{renorm-dual}) means that the vacuum of 
electrically charged particles shields the external electromagnetic field 
but the contribution from magnetically charged particles antishields it.
This agrees with the results of \cite{Giebl} and \cite{TTS}.

Considering now the case of weak electromagnetic fields,
the finite part of the Lagrangian $\Delta L'$, can, by analogy with
(\ref{DeltaL-Euler}), be written as
\begin{eqnarray}                           \label{DeltaL-dyon}
\Delta L'
& = & 
\frac{1}{360 \pi ^2 M^4} 
\bigl\{ [ (Q^2 - g^2)^2  + 7 Q^2g^2] (\vecH^2 - \vecE^2)^2 
+ [16 Q^2 g^2 + 7 (Q^2 - g^2)^2](\vecH \vecE)^2
\nonumber  \\
& &\hspace*{20mm}
+ 6 Qg (Q^2 - g^2)(\vecH \vecE)(\vecH^2 - \vecE^2) \bigr\}.
\end{eqnarray}

The expressions (\ref{DeltaL-Euler}) and (\ref{DeltaL-dyon}) 
describe nonlinear corrections to the Maxwell equations 
which correspond to photon-photon interactions.
The principal difference between the formula
(\ref{DeltaL-dyon}) and the standard Euler-Heisenberg effective Lagrangian
consists in the appearance of $P$ and $T$ non-invariant terms 
proportional to $(\vecH \vecE)(\vecH^2 - \vecE^2)$. 
It should however be noted that this term is invariant under
charge conjugation $C$, since then {\it both} $Q$ and $g$ would
change sign.

Thus, the matrix element of the photon-photon interaction will contain
terms which violate $P$ and $T$.

If we consider separately the virtual creation of dyon pairs,
then because of invariance of the model under a dual
transformation  (see, e.g. \cite{StrTom}) the physics is determined
not by the values $Q$ and $g$ separately, but by the effective charge
$\sqrt{Q^2 + g^2}$. 
In the same way the operations of $P$ and $T$ inversions are modified.
However we will consider simultaneously the contributions from vacuum
polarization by electron-positron and dyon pairs.
In this case it is not possible to reformulate the theory in terms of 
just one effective charge by means of a dual transformation. 
Moreover the Dirac charge quantization condition
connects just the electric charge of the electron 
and the magnetic charge of
a dyon: $eg = n/2$ whereas the electric charge $Q$ is not quantized.

It is widely believed, based both on 
experimental bounds and theoretical predictions \cite{GodOlive}
that the dyon mass would be large, $M \gg m$, where $m$ is the electron
mass.
Thus, in the one-loop approximation the first non-linear correction 
to the QED Lagrangian from summing the contributions
(\ref{DeltaL-Euler}) and (\ref{DeltaL-dyon}) can be written as
\begin{equation}                            \label{DeltaL-dyon-e}
\Delta L' \approx
\frac{e^4}{360 \pi ^2 m^4} \bigl[
(\vecH^2 - \vecE^2)^2 +  7 (\vecH \vecE)^2 \bigr]
+ \frac{Qg (Q^2 - g^2)}{60 \pi ^2 M^4} 
(\vecH \vecE) (\vecH^2 - \vecE^2),
\end{equation}
where the $P$ and $T$ invariant terms corresponding to vacuum polarization
by dyons have been dropped because they are suppressed by factors
$M^{-4}$. 
Thus, their contribution to the effective Lagrangian will be
of the same order as that of the ordinary QED multiloop amplitudes 
which we neglect.

This expression (\ref{DeltaL-dyon-e}) for the effective Lagrangian 
allows us to calculate the amplitude for photon splitting in an external
non-spatially uniform magnetic field $\vecH$
\begin{equation}\gamma (k) + ~{\mbox {\rm external magnetic field }\vecH}
\to \gamma (k_1) + \gamma (k_2) .
\end{equation}

In order to determine this amplitude, we shall follow the approach 
by \cite{Adler} (see also \cite{BerLifPit}). 
We may write the matrix element of the process in terms of
functional derivatives of the effective Lagrangian:
\begin{align}                                   \label{Matr}
M 
& = 
\delta E^i \delta E^j_{1}\delta E^k_{2}
\frac{\delta^3 L}{\delta E^i \delta E^j \delta E^k}
\bigg|_{\Atop{\vecE = 0}{|\vecH| = H}}  +
\delta H^i \delta H^j_{1} \delta H^k_{2}
\frac{\delta^3 L}{\delta H^i \delta H^j \delta H^k}
\bigg|_{\Atop{\vecE = 0}{|\vecH| = H}}
\nonumber \\
&
+   (\delta E^i \delta E^j_{1} \delta H^k_{2} 
+    \delta E^i_{1} \delta E^j_{2} \delta H^k
+    \delta E^i_{2} \delta E^j \delta H^k_{1})
\frac{\delta^3 L}{\delta E^i \delta E^j \delta H^k}
\bigg|_{\Atop{\vecE = 0}{|\vecH| = H}}
\nonumber \\
&
+   (\delta H^i \delta H^j_{1} \delta E^k_{2}
+    \delta H^i_{1} \delta H^j_{2} \delta E^k
+    \delta H^i_{2} \delta H^j \delta E^k_{1})
\frac{\delta^3 L}{\delta H^i \delta H^j \delta E^k}
\bigg|_{\Atop{\vecE = 0}{|\vecH| = H}}
\end{align}
where the subscripts 1 and 2 label the photons in the final state.
So, the photons are just one-particle fluctuations over the considered
vacuum field configuration, i.e.,
$\vecE = 0 + \delta \vecE$; 
$\vecH = \vecH_0 + \delta \vecH$,
and the photon field strengths are
\begin{align}
\delta H & = \sqrt{4\pi}\, \omega (\hat \veck
\times \hat \vecvarepsilon), & 
\delta E & = \sqrt{4\pi}\, \omega \hat \vecvarepsilon, \nonumber  \\
\delta H_{1} & = \sqrt{4\pi}\, \omega_1 (\hat\veck_1
\times \hat\vecvarepsilon_1), & 
\delta E_1 & = \sqrt{4\pi}\, \omega_1 \hat \vecvarepsilon_1, 
\nonumber  \\
\delta H_{2} & = \sqrt{4\pi}\, \omega_2 (\hat \veck_2
\times \hat\vecvarepsilon_2), & 
\delta E_2 & = \sqrt{4\pi}\, \omega_2 \hat \vecvarepsilon_2,
\end{align}
Here we have not written out phase factors which are common to all terms 
and the unit vectors of the respective photon polarizations and momenta
are $\hat\vecvarepsilon$, $\hat\vecvarepsilon_1$, 
$\hat\vecvarepsilon_2$ and $\hat \veck$, $\hat \veck_1$, $\hat 
\veck_2$.
For the case of small spatial variations of the external
magnetic field, the change $\vecp$ of the total momentum 
would be rather small. The condition becomes \cite{Adler}
\begin{equation}
\omega = \omega_1 + \omega_2,
\qquad \vecp  + \veck
= \veck_1 + \veck_2
\end{equation}
\noindent
Thus to the first order in the parameter $(|\vecp|/\omega)^{1/2}$
the angles between the photon momenta, given by
$\cos\phi_1=\hat \veck\cdot\hat \veck_1$,
$\cos\phi_2=\hat \veck\cdot\hat \veck_2$,
$\cos\phi_{12}=\hat \veck_1\cdot\hat \veck_2$,
can be approximated as \cite{Adler}
\begin{align}                                \label{phis}
\phi _1 
&\approx
\left(\frac{\omega_2}{\omega_1}\right)^{1/2}
\left( -2 \frac{p}{\omega} {\hat \vecp} \cdot {\hat \veck} 
\right)^{1/2},
\nonumber  \\
\phi _2 
&\approx
\left(\frac{\omega_1}{\omega_2}\right)^{1/2}
\left( -2 \frac{p}{\omega} {\hat \vecp} \cdot {\hat \veck} 
\right)^{1/2},
\nonumber  \\
\phi _{12} 
&\approx
\frac{\omega}{(\omega_1 \omega_2)^{1/2}}
\left( -2 \frac{p}{\omega} {\hat \vecp} \cdot {\hat \veck} 
\right)^{1/2}
= \phi _1 + \phi _2,
\end{align}
where use has been made of the fact that the final-state
photons must lie in the plane defined by the magnetic field
and the incident photon momentum.

The probability for photon splitting in an external field
depends crucially on the polarizations of the photons
\cite{Adler}, \cite{BerLifPit}.
In the case of small-angle scattering, only the process
$\gamma_{\|} \to \gamma_{\bot1} + \gamma_{\bot2} $
takes place. Here the indices $\|$ and $\bot$
correspond to polarization states where the direction 
of the magnetic field of the photon
$\vecH_i = \sqrt{4\pi}\, \omega ({\hat \veck} 
\times \hat \vecvarepsilon)$ is either perpendicular to 
or parallel to the plane formed by 
the momentum of the initial photon $\veck$ and 
the external magnetic field $\vecH$.

Thus after substitution of Eq.~(\ref{DeltaL-dyon-e}) into (\ref{Matr}) 
and some lengthy but rather simple calculations 
we arrive at the expression for the matrix element for
photon splitting in an external magnetic field, 
taking into account the extra term which corresponds to 
the dyon-loop contribution:
\begin{equation}                            \label{Matr-final}
M_{\gamma\to2\gamma} = -\frac{4}{15\sqrt{\pi}}
\biggl( \frac{e^4}{3 m^4} \sin \theta
\left[\frac{7}{4} (\phi_1^2 + \phi_2^2) - \phi_{12}^2\right]
-\frac{Qg (Q^2 - g^2)}{8 M^4} \cos \theta\,
\phi _{12}^2 \biggr) 
\omega \omega_1 \omega _2  H \cos 2 \beta
\end{equation}
where $\theta$ is the angle between the momentum $\veck$ 
and the magnetic field $\vecH$,
$\beta$ the dihedral 
angle between the plane containing 
the above vectors and the one containing the momenta of the
final-state photons, and the $\phi$'s given by Eq.~(\ref{phis}).

As was noted above, the most important difference 
between Eq.~(\ref{Matr-final})
and the standard formula for the photon splitting matrix element
\cite{Adler} consists in the appearance of 
an additional $P$ and $T$ non-invariant term.
Indeed, for the crossed process of two photons coalescing
in an external field 
 \begin{equation}
\gamma (k_1) + \gamma (k_2) \to
\gamma (k) + ~{\mbox {\rm external magnetic field H}}~
\end{equation}
the  
matrix element $M_{2\gamma \to \gamma}$ can be just obtained
from Eq. (\ref{Matr-final}) by reflection.
Thus, the second term in (\ref{Matr-final}) under this operation
has to change its sign and we have 
\begin{equation}                                \label{Mat-final}
M_{2\gamma \to \gamma} = 
 -\frac {4}{15\sqrt{\pi}} \biggl( \frac{e^4} {3  m^4} 
\sin \theta
\left[\frac{7}{4} (\phi _1^2 + \phi_2^2) - \phi _{12}^2\right] 
 +
\frac{Qg (Q^2 - g^2)}{8 M^4} \cos \theta\,
\phi _{12}^2 
\biggr) \omega \omega_1 \omega _2  H \cos 2 \beta
\end{equation}

Thus, as a result of interference between two one-loop diagrams 
corresponding to loops with dyons and those with simply electrically 
charged particles there is an asymmetry between 
the processes of photon splitting and photon coalescence. 
The physical effect of this asymmetry will depend on the photon
spectrum and the directions of the photon momenta with respect 
to the magnetic field.  In particular, the asymmetry vanishes 
when these are perpendicular, i.e. for $\cos\theta=0$.
A simple measure is the factor of asymmetry \cite{Khriplovich},
which for values of $\theta$ not too close to zero takes the form
\begin{equation}
\delta = \frac 
{|M_{2\gamma \to \gamma}|^2 - |M_{\gamma \to 2\gamma}|^2}{  
 { |M_{2\gamma \to \gamma}|^2 + |M_{\gamma \to 2\gamma}|^2}}
\simeq {\frac{3 Q g (Q^2 - g^2)}{e^4}}~ {\frac {m^4}{M^4}}~ 
{\frac{\phi _{12}^2}{7 (\phi _1^2 + \phi _2 ^2) - 4 \phi _{12}^2}}
 \cot\theta, \qquad \theta\ge {\cal O}\left(\frac{m}{M}\right)^4
\end{equation}
which, as expected, is linear in the product of the dyon charges, 
and proportional to the fourth power of the electron to dyon mass ratio.
Possible consequences of this effect in the context of 
Early Universe evolution will be discussed elsewhere.

\bigskip

{\bf Acknowledgements}
\medskip

We acknowledge numerous conversations with A. Gazizov and V. Kiselev.
One of us (Ya. S.) is very indebted to Prof.\ I.B. Khriplovich for
fruitful discussions. 
He would also like to thank Prof.\ G. Calucci for
very helpful remarks and  
the International
Atomic Energy Agency and UNESCO, for hospitality at International
Centre for Theoretical Physics, Trieste, where this work has been
completed.
Ya.~S. also acknowledges on the first stage of work support by the 
Fundamental Research Foundation of Belarus, grant No F-094.
This research (P. O.) has been supported by 
the Research Council of Norway.

\end{document}